\documentclass[]{emulateapj}
\usepackage{graphicx} 
\usepackage{ulem}
\usepackage{rotating}
\usepackage{lscape}

\newcommand{\kms}{km~s$^{-1}$ }
\newcommand{\cm}[1]{\, {\rm cm^{#1}}}

\newcommand{\delv}{\Delta v}

\newcommand{\sci}[1]{{\rm \; \times \; 10^{#1}}}

\begin{document}

\submitted{Accepted for publication in The Astrophysical Journal}
\title{Spatially Correlated Cluster Populations in the Outer Disk of NGC 3184}
\author{St\'ephane Herbert-Fort$^1$, Dennis Zaritsky$^1$, John Moustakas$^2$,\\
Daniel Christlein$^3$,  Eric Wilcots$^4$, Andrea Baruffolo$^5$,\\
Andrea DiPaola$^6$, Adriano Fontana$^6$, Emanuele Giallongo$^6$,\\
Richard W.\ Pogge$^7$, Roberto Ragazzoni$^5$, Riccardo Smareglia$^8$}
\vspace{0.15cm}
\affil{
$^1$University of Arizona/Steward Observatory, 933 N Cherry Avenue, Tucson, AZ 85721\\
 (email: shf@as.arizona.edu) \\
$^2$Center for Cosmology and Particle Physics, New York University, 4 Washington Place, New York, NY 10003\\
$^3$Max-Planck-Institut f\"ur Astrophysik,
Karl-Schwarzschild-Str. 1, 85748 Garching, Germany\\
$^4$Department of Astronomy, University of Wisconsin-Madison, 475 N. Charter St., Madison, WI 53706\\
$^5$INAF, Osservatorio Astronomico di Padova, vicolo dell'Osservatorio 5, I-35122 Padova, Italy\\
$^6$INAF, Osservatorio Astronomico di Roma, via di Frascati 33, I-00040 Monteporzio, Italy\\
$^7$Department of Astronomy, The Ohio State University, 140 W. 18th
Avenue, Columbus, OH 43210-1173\\
$^8$INAF, Osservatorio Astronomico di Trieste, via G. B. Tiepolo 11, I-34131 Trieste, Italy
\vspace{0.05cm}
}

\begin{abstract}

We use deep ($\sim27.5$ mag $V$-band point-source limiting magnitude) $V$-\ and $U$-band LBT imaging to 
study the outer disk (beyond the optical radius $R_{25}$) of the non-interacting, 
face-on spiral galaxy NGC 3184 ($D = 11.1$ Mpc; $R_{25} = 11.1$ kpc) 
and find that this outer disk contains 
$>1000$ objects (or marginally-resolved `knots') resembling star clusters 
with masses $\sim10^2 - 10^4 M_{\sun}$ and ages up to $\sim1$ Gyr. 
We find statistically 
significant numbers of these cluster-like knots 
extending to $\sim1.4 R_{25}$, with the redder knots outnumbering bluer at 
the largest radii.  
We measure clustering among knots and find 
significant correlation to galactocentric radii of $1.5 R_{25}$ for 
knot separations $<1$ kpc.  
The effective integrated surface brightness of this outer disk 
cluster population ranges from $30 - 32$ mag arcsec$^{-2}$ in $V$.  
We compare the \ion{H}{1} extent to that of the correlated knots 
and find that the clusters extend at least to 
the damped Lyman-$\alpha$ threshold of \ion{H}{1} column density ($2\sci{20} \cm{-2}$; $1.62 R_{25}$).  The blue knots are correlated with \ion{H}{1} spiral 
structure to $1.5R_{25}$, while the red knots may be correlated with 
the outer fringes of the \ion{H}{1} disk to $1.7R_{25}$.  These results 
suggest that outer disks are well-populated, common, and long-lasting 
features of many nearby disk galaxies.
\end{abstract}


\section{Introduction}

Ultraviolet observations with the Galaxy Evolution Explorer 
\citep[GALEX;][]{Martin05} indicate that $\sim30\%$ of 
nearby disk galaxies host star formation in an extended component 
\citep{Thilker07,Zaritsky07}.  These stellar disks sometimes reach 
to more than twice the optical radius ($R_{25}$) and challenge 
our understanding of disk galaxies.  
Motivated by a lack of understanding and because progress here may 
pose additional challenges for galaxy formation models, 
we begin a systematic investigation of a sample of nine nearby 
galaxy outer disks using deep optical imaging and present first results 
here. 

Although deep H$\alpha$ and broadband optical observations 
\citep[e.g.][]{Ferguson98, Martin01, Weiner01} 
had previously detected star formation in outer disks,
the ubiquity, and therefore the significance, of this component 
has been largely overlooked until recently.  The traditional tracer of 
very young star clusters and ongoing star formation, H$\alpha$ flux, is emitted by 
gas surrounding a cluster for a short period of time 
($\sim10$ Myr, or about the lifetime of massive OB stars).  
Consequently, the number of H$\alpha$ detections is modest 
\citep{Ferguson98, Martin01, Weiner01}.  

The two most prominent extended disks studied with GALEX so far, M83 and NGC 4625 
\citep{Thilker05, GildePaz05}, have nearby companions, suggesting that gravitational
interactions may be responsible for their extended outer disks.
A more comprehensive GALEX study \citep{GildePaz07} 
does contain isolated galaxies, but typically they are 
more distant (the majority of galaxies in the GALEX Atlas sample are $>20$ Mpc away) 
and so only the brightest end of the knot 
luminosity function is sampled.
To address these shortcomings, we select nearby, isolated galaxies, observe them sufficiently 
deeply to probe the knot luminosity function well and 
obtain numerous candidate knots, and develop a method to quantify the nature 
of the knot spatial distribution.

In complement to the imaging studies, \cite{Christlein08} have shown that 
deep longslit spectroscopy can provide a measure of 
both the integrated extent and global disk-plane 
kinematics (i.e.\ rotation curves) of outer disks in edge-on galaxies.  
While spectroscopy provides useful information, it is generally 
limited to knots emitting H$\alpha$.  
To expand the lookback time over which we can study disks, we 
image at redder colors to identify older candidate knots, but 
the drawback of this approach is that discrimination with background 
sources becomes more challenging. 

Because we use optical observations to search for outer disk knots analogous to those detected 
by GALEX, we note the differences between the two sets of data.  
First, GALEX's spatial resolution ($\sim5''$ FWHM) is roughly 
six times larger than what we typically achieve.  At the distance of NGC 3184, 
GALEX's resolution element corresponds to a physical scale of $\sim270$ pc, 
compared to the $\sim40$ pc of our ground-based data.  
GALEX knots are typically blends of multiple clusters \citep{GildePaz05}, 
whereas (as we will show), we detect knots resembling individual star clusters.  
Second, GALEX observations are sensitive mainly to young clusters 
\citep[$<500$ Myr;][]{Thilker05} whereas we are sensitive to 
clusters with ages up to several gigayears.  Finally, our mass limit is 
lower by a factor of 10. 
Our LBT data will therefore provide 
larger numbers of knots to use in any statistical measure of the disk.

We present the initial results of a statistical study of nearby ($<15$ Mpc) outer disks, using 
the 8.4m Large Binocular Telescope \citep[LBT, Mt.\ Graham, Arizona;][]{Hill06} 
and wide-field, prime-focus Large Binocular Cameras \citep[LBC;][]{Ragazzoni06,Giallongo08}.  
We describe our data reduction, develop analysis tools, and apply these to 
deep $V$-band and $U$-band (hereafter $V$ and $U$) imaging data 
of the nearly face-on \citep[inclination $= 17^\circ$;][]{Daigle06} 
spiral galaxy NGC 3184 \citep[$D = 11.1$ Mpc, or $m-M = 30.23$ mag;][]{Leonard02}. 
NGC 3184 is similar to an $L$* galaxy, with 
$M_V = -20.8 $ and $M_* \sim 1.4\sci{10} M_{\sun}$ \citep{Moustakas09}.
We demonstrate how two separate statistical methods enable us to trace outer 
disk cluster-like objects to large radii.  
Because the distance to NGC 3184 is uncertain at the $\sim20\%$ level ($\sim2$ Mpc), 
all scales referenced to the frame of the disk are also uncertain 
at the $\sim20\%$ level.  
This uncertainty in distance does not significantly affect our results.  
The detected correlation signals are independent of 
the physical size of the field.  

We address the following questions here:  1) In a non-interacting galaxy,
without an obvious population of GALEX knots,  
do we detect any evidence of an extended disk?  2) If so, what are the 
oldest clusters we detect?  3) How can we detect an older population
of knots that is less distinct from the background galaxy
population?  
4) How does the extent of the disk compare to the fuel source (i.e.\ the gas).

In \S2 we describe our observations, data reductions and source detections.  In \S3 we present 
color-magnitude diagrams (CMDs) of candidate outer disk sources and 
the range of cluster properties consistent with the candidate knots.  We also present an estimate 
of disk knot radial extent determined from the CMDs in \S3.  
In \S4 we present a restricted three-point correlation analysis that increases 
contrast with the background and more effectively probes the extent of clustered 
outer disk objects.  In \S5 we present a similar clustering analysis 
using GALEX UV sources, and in \S6 we present a comparison to the underlying neutral 
gas profile, using 21cm VLA data with a $3\sigma$ detection limit of $N(\rm{HI}) = 6.9\sci{19} \cm{-2}$.  
In \S7 we discuss the local environment of NGC 3184.  
We present a summary of our results and a brief discussion in \S8.

\section{Observations to Final Source Catalog}

\subsection{Observations}

We observed NGC 3184 with LBC-Blue on the LBT during Science 
Demonstration Time on March 20, 2007.  We obtained 
eight and nine dithered, 164-second exposures through the $V$ and $U$ filters, 
respectively, under non-photometric conditions and $\sim0\farcs8$ seeing, and 
obtained single 164-second photometric exposures (in $V$ and $U$) on 
February 2, 2008, together with three photometric Landolt 
standard star fields \citep{Landolt92} for flux calibration.  The photometric exposures served their purpose 
but due to poor image quality we exclude them from the final science mosaics.

\subsection{Data Reduction}

We correct for both global changes and the two-dimensional structure 
in the bias using a set of Interactive Data Language (IDL\footnote{developed by Research Systems, Inc.\ and 
owned by ITT; http://www.ittvis.com/ProductServices/IDL.aspx}) scripts created by our group.
The LBC-Blue detector array consists of four 2k x 4k CCD chips 
(each with a gain of $\sim1.75$ electrons/ADU, read noise 
of $\sim12$ electrons and $\sim0\farcs22$-wide pixels), three aligned side-by-side 
lengthwise and one centered perpendicularly above them.  We begin by correcting for 
bias gradients along columns (spanning the long axis) in the bias frames.  
The median within the overscan region along each row 
is subtracted from that row.  
The overscan-corrected bias frames of each chip are then median combined 
after removing 1$\sigma$ outliers, where $\sigma$ is calculated excluding the 
minimum and maximum values in the stack (15 frames/stack were used).  
We adopt an aggressive clipping (1$\sigma$) to ensure that deviant values, 
which artificially increase the calculated rms value, are excluded.  
The result is our final `master bias' frames used 
to subtract any residual structure in an image after the overscan levels 
are accounted for.  We bias-correct all raw frames in this manner and trim 
the overscan regions.

To correct for sensitivity variations on the CCDs, we combine dithered, twilight-sky flat-field 
images (10 in each band).  We calculate the four-chip median value of each bias-corrected 
flat field exposure and divide each chip 
of the particular four by this median level to normalize the flat.  
Optical distortions in LBC-Blue can result in non-flat images (a $\sim5\%$ effect 
across the field of view\footnote{http://lbc.oa-roma.inaf.it/commissioning/flatfield.html}).  
We correct for this effect later  
using the distortion maps created by SCAMP\footnote{Version 1.4.0; http://terapix.iap.fr/soft/scamp} 
\citep{Bertin06} 
when aligning our individual images to a common field of view (see below).  
The normalized flats are then median combined after 
minimum/maximum and $1\sigma$ rejection as above, producing our 
`master flat' for each chip.  
We complete our processing by dividing the bias-corrected science frames 
by the master flats, resulting in $V$ images flat to $\sim 0.5\%$ 
and $U$ images flat to $\sim 1\%$.  

We next subtract the background level.  
Estimating the true background is made difficult by 
contaminating scattered light that varies sensitively with the telescope orientation, 
and by the extended 
galaxy that covers a large portion of the field in each exposure.  
To minimize the effects of these contaminants, 
we determine the background level in 
twenty $200 \times 200$ pixel regions distributed near the edges of the detector array 
using the IDL routine \textsc{mmm}\footnote{part of the Goddard IDL library, maintained by 
W.\ Landsman; http://idlastro.gsfc.nasa.gov/}, which estimates the sky background in a 
stellar contaminated field by assuming that contaminated sky pixel values 
overwhelmingly display positive departures from the true value.  We then adopt the 
minimum background value from the regions on each chip and subtract it 
from that particular chip.  This process is done separately for 
each exposure.

To create combined images free of spurious signal or cosmetic defects, 
we create masks of bad columns and hot pixels for each chip interactively 
(based on obvious defects in the flats) and 
invert the masks to create the weightmaps used 
in the combining process.  
Next, we reject cosmic ray detections with the IDLUTILS\footnote{http://spectro.princeton.edu/idlutils\_doc.html} 
routine \textsc{reject\_cr} 
(written by M.\ Blanton), which rejects cosmic rays by 
finding features sharper than the point spread function (PSF) at $>6\sigma$ above 
the background.  
The identified cosmic ray pixels are incorporated into the bad pixel masks.

\subsection{Mosaic Creation}

We produce our final mosaics using SCAMP and SWarp\footnote{Version 2.17.1; http://terapix.iap.fr/soft/swarp}.  
SCAMP solves the astrometry of the dithered exposures, and requires that we first 
create catalogs of detected sources in each exposure.  
To create the necessary catalogs for SCAMP, we use 
SExtractor \citep{Bertin96}, restricting source detection to 
objects with five or more adjacent pixels for which the pixel values 
are at least 8$\sigma$ above sky in $V$ and 3$\sigma$ above sky in $U$.  
We exclude pixels with values above 40,000 ADU when running SExtractor 
because these pixels produced false detections ($\sim65000$ ADU is the raw 
saturation level of the CCDs).  
We find that the above choices result in the best distortion 
maps from SCAMP (maps 
whose distortion level contours trace smoothly across the chip gaps), 
using three SCAMP iterations with tightening criteria for source matching.  
We use the SDSS fifth data release \citep[DR5;][]{DR5} as the reference catalog 
to match sources in the NGC 3184 field.  
SCAMP provides the WCS solutions used for combining the frames with 
SWarp.  
We set the `EXPOTIME\_KEY' parameter in SCAMP to `DONTUSE', so that the 
resulting fluxes are not altered by the image exposure times 
(we normalize the photometry to a fixed second$^{-1}$ standard after 
combining the images with SWarp, when calibrating the science images to the 
standard star images).  SCAMP is used to create the distortion maps that 
correct for the optical concentration effect mentioned earlier, so that the 
images are on the same $\sim0\farcs22$ pixel$^{-1}$ spatial scale before 
being combined into a mosaic. 

The final step is to create the deep mosaics we use for source detection, 
calibration and analysis.  We pass the output headers from SCAMP containing 
the WCS solutions of each exposure to SWarp, which we set to combine the 
spatially-aligned frames to their average values after accounting for 
the bad pixel masks.  
Total integration times in the deepest regions of the resulting mosaics are 
$\sim 25$ and $\sim 22$ minutes in $V$ and $U$, respectively. 

\subsection{Source Detection}

We build final $V$ and $U$ aperture photometry catalogs using SExtractor, 
with source detection and aperture placement 
based on the deeper $V$-band mosaic (both mosaics have the same $\sim0\farcs8$ PSF, 
and similar image quality).  
After trying various combinations of parameters, 
we select sources by identifying groups of five or more pixels each with 
flux $>1\sigma$ above the background (so typical `detections' are 
actually $>3\sigma$, because $[<0.3173]^{[\geq5]} < 0.0032$).  
For our parameter choices, we set the saturation level to be 
55,000 counts to conservatively avoid 
problematic saturated areas (we are less concerned with false detections in 
the mosaic, particularly with the fine-tuned settings listed below), 
a seeing FWHM of $0\farcs8$, a pixel 
scale of $0\farcs22$, local background calculation, the background manually set as 0.0 
because we have already subtracted it (note that the photometry uses 
the local background, however), a 3x3-pixel `all-ground' convolution mask with 
FWHM=2 pixels for filtering to avoid spurious 
detections of noise peaks, a minimum contrast deblending parameter of 0.001, a 
`cleaning' of the catalog with an efficiency parameter of 1, a background mesh size of 
32 pixels, and a background-filtering mask size of 3 background meshes. 
Using the same SExtractor parameters, we then 
perform matched-aperture photometry on the $U$ image.  
A visual inspection of the detected sources 
confirms that the above parameters lead SExtractor to 
detect nearly all visually-discernable objects 
in the region of interest, beyond the optical radius $R_{25}$.  
Our catalog becomes noticeably incomplete below $\sim0.8R_{25}$, where SExtractor has 
difficulty detecting unique sources over the extended, bright emission of the inner disk.

\begin{figure}[h]
\centerline{
\includegraphics[angle=0,width=3.5in]{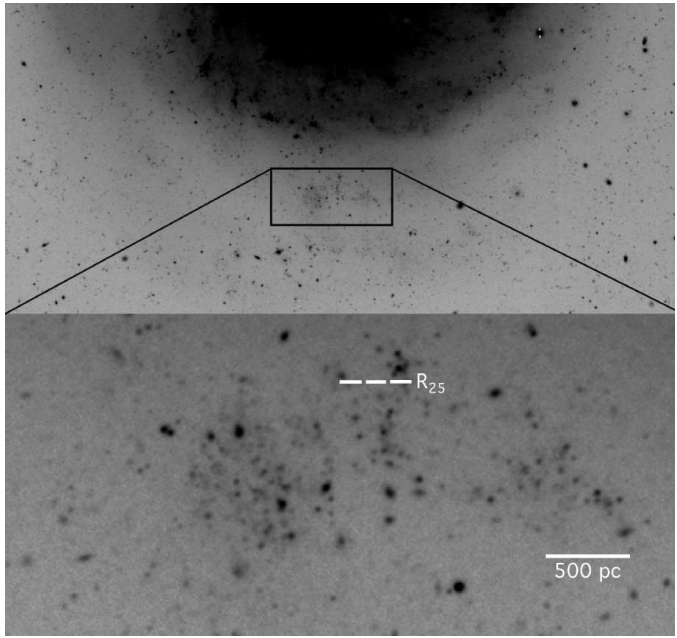}}
\caption{A portion of our deep $V$-band mosaic, with a zoom-in on a complex of 
faint knots near the 
optical radius.  The optical radius ($R_{25}$) and a 500 pc scale bar are shown in 
white in the lower image.  Many of the knots are 
marginally-resolved, non-uniform sources.  
The majority of sources in this zoomed-in view have colors far bluer 
than the typical background galaxy and are grouped in a manner 
to suggest that they are in fact disk objects.  The physical scales of the individual 
knots is $\sim40$ pc at the distance of NGC 3184.
\label{knots_fig}}
\end{figure}

Given the distance of NGC 3184 (11.1 Mpc, or $m-M = 30.23$ mag), our $0\farcs8$ 
seeing ($\sim40$ pc at the distance of NGC 3184), and that all but the brightest OB stars 
would fall below our detection limit ($\sim27.5$ mag.\ for the bluest sources), 
the candidate outer disk objects are likely to be groupings of stars.  
Because we detect the integrated light 
from members of stellar groups, or knots (see Figure~\ref{knots_fig} for examples), 
any photometric algorithm that requires a uniform 
object shape for extraction is not optimal for this work.  

\subsection{Photometry \& Final Source Catalog}

We calibrate our photometry using photometric exposures of Landolt standard 
star fields, taken on the same night as our individual photometric 
exposures of NGC 3184, to flux-calibrate the observed standard star signal 
for a range of airmasses and colors.  
We then place the photometry
of eight isolated stars in the photometric frame of NGC 3184 on the standard Vega system, 
accounting for the calculated airmass and color terms as well 
as the different exposure times, and finally 
bootstrap the photometry of the deep mosaic using these same stars.

Colors and magnitudes are measured using circular apertures.  
Colors quoted throughout are from fixed 
four-pixel-diameter apertures ($0\farcs9$ or $\sim48$ pc at NGC 3184, just larger than the 
typical $0\farcs8$ FWHM of detected sources), 
while $V$ magnitudes are from 10-pixel-diameter apertures 
($2\farcs24$ or $\sim121$ pc at NGC 3184) 
and aperture-corrected using stellar curves-of-growth 
by $-0.08$ mag.  Aperture corrections were calculated from eight isolated, 
unsaturated stars measured in 15 apertures 
spanning $2-50$ pixels in diameter (or $0\farcs4 - 11\farcs2$).  
To the degree the knots are resolved, 
our aperture corrections will underestimate the total flux.  
We choose the 10-pixel-diameter apertures so as to include the bulk flux from the many 
marginally-resolved extended sources visible in the images while 
not including too many neighbors (see Figure~\ref{knots_fig}; false-color composite images are 
available from the electronic version of the paper).   
We set SExtractor to mask and correct neighbors that contaminate.  

In our final catalog, we only include sources 
whose $U-V$ color error is $<0.5$ mag (magnitude errors are provided 
by SExtractor and propagated in the standard manner).  
This leaves $\sim4500$ sources between $1.0-1.5 R_{25}$, 
the `outer disk region' we examine below (the outer limit, $1.5 R_{25}$, is an 
arbitrary choice here).

As a check of our photometry, we compare the 
apparent magnitudes of ten well-isolated objects across the field with those provided by 
SDSS-DR5, converted from $u$, $g$, and $r$ to either $V$ or $U$ using the 
transformations of \cite{Jester05}.  We find that our results are consistent 
with the transformed SDSS photometry to within the transformed 
SDSS and LBT photometric errors.

\section{Cluster Populations Surrounding NGC 3184}

\begin{figure}[h]
\vspace{0.2cm}
\centerline{
\includegraphics[angle=90,width=3.5in]{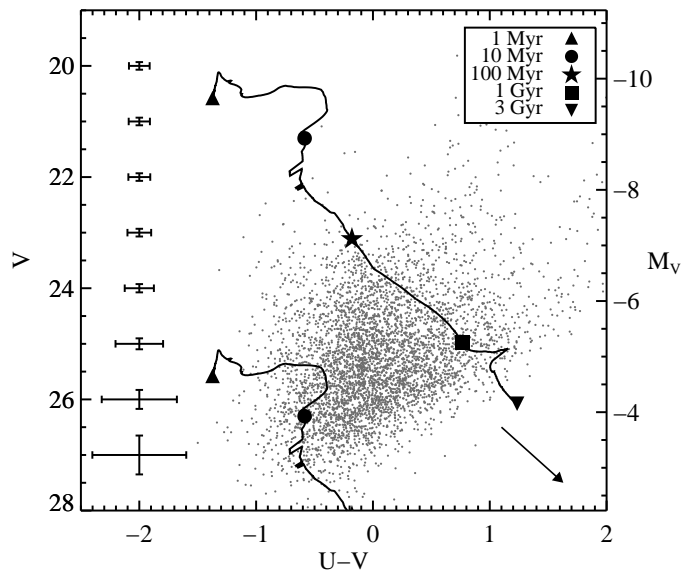}}
\caption{CMD of $\sim4500$ sources between $1.0 - 1.5 R_{25}$ 
from our final catalog, with 
median 1$\sigma$ errors descending the plot at left.  
A reddening vector corresponding to 1 magnitude of extinction in $V$, calculated 
using results from \cite{Rieke85}, is shown at lower right.  See the text for a 
description of the Starburst99 model tracks. 
\label{colmag}}
\end{figure}

Figure~\ref{colmag} shows the CMD of all detected 
sources between $1.0 - 1.5 R_{25}$ ($R_{25} \sim3.45$ arcmin, or $\sim11.1$ kpc).  
Overplotted in black are Starburst99~\citep{Leitherer99,Vazquez05} models of 
fixed mass, solar metallicity star clusters covering a range in their 
evolutionary sequence from 1 Myr to 3 Gyr.  
The upper and lower tracks are of $10^4 M_{\sun}$ and 
$10^2 M_{\sun}$ model clusters, respectively, 
scaled down in mass/magnitude from a simulated $10^6 M_{\sun}$ cluster that 
adequately samples the Kroupa IMF.  The scaled tracks are meant as a general 
guide only; the stochastic sampling of the IMF at low cluster masses \citep{Cervino04,Fagiolini07} 
is not accounted for in the following knot mass and age estimates.  The uncertainties 
from the stochastic sampling are larger for the lower-mass and younger clusters, and a scaled-down model cluster track 
(from a well-sampled IMF) becomes systematically brighter and bluer than real 
clusters would be at a particular age (the tracks represent contributions 
from the highest-mass stars, which are increasingly unlikely to be found in real low-mass clusters).  
As a result, comparisons with the scaled model tracks can lead to underestimates of cluster masses 
and overestimates of their ages.  Our aim here is to provide a general impression only of the cluster 
masses and ages consistent with the knots in our sample.

To statistically constrain the color and magnitude range 
of sources most likely associated with NGC 3184, we create 
background-subtracted Hess diagrams.  A Hess diagram plots the 
number of sources within chosen color-magnitude bins across a CMD.  
The extent of the remaining signal in the background-subtracted 
diagrams provides an estimate of the size of the outer disk.  
To create these diagrams, we first produce a `background' CMD 
from sources detected in an outer annulus between $2 - 3\ R_{25}$.  We also create 
a `disk+background' CMD from the region of interest and then 
count the number of sources found in bins of color and 
magnitude (square bins of 0.2 mag were used here).
We then scale the background Hess diagram to match 
the area on the sky represented by the disk+background Hess diagram.  
Finally, we subtract the scaled 
background Hess diagram from the 
disk+background Hess diagram to reveal the signal above the background.

\begin{figure*}
\centerline{
\includegraphics[angle=90,width=6.0in]{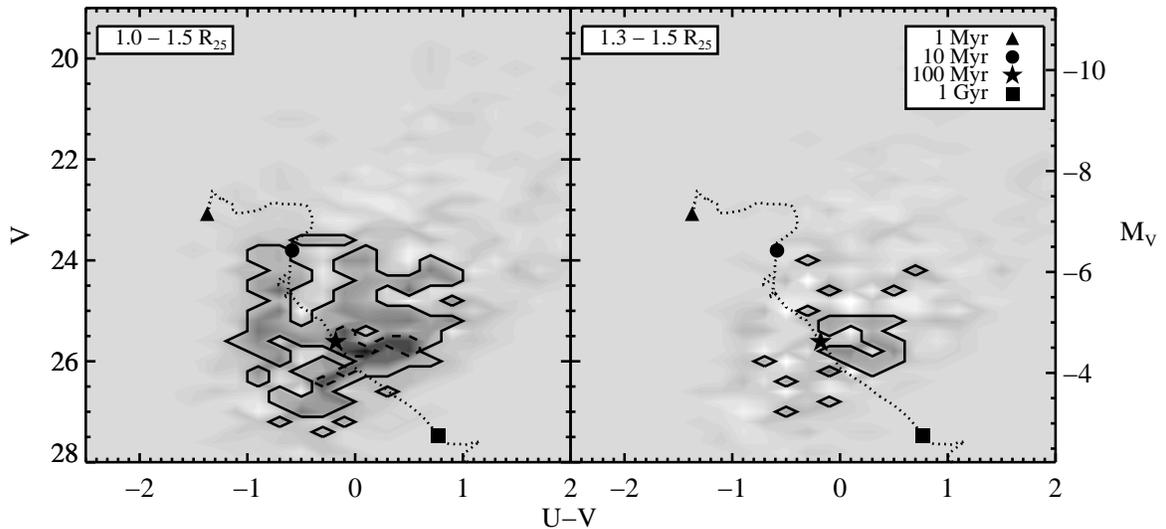}}
\caption{Background-subtracted Hess diagram showing the number of 
sources remaining between $1.0 - 1.5\ R_{25}$ (left panel; dark regions are positive counts) 
and between $1.3 - 1.5\ R_{25}$ (right panel).
Solid and dashed black 
contours outline signal lying above the background at $>90\%$ and $>99.9\%$ CL, 
respectively.  Overplotted is a $10^3 M_{\sun}$ Starburst99 model cluster, scaled down from a $10^6 M_{\sun}$
cluster as in Figure~\ref{colmag}.  {\it{Left:}} this `outer disk CMD' shows a faint blue plume of sources 
near $U-V = -1$.  The strongest signal arises from redder sources 
at faint levels (V $\sim26$ mag.).  
From the original $\sim4500$ sources in Figure~\ref{colmag},
$\sim1100$ (or $\sim1/4$) remain after accounting for the scaled background CMD; taking the 
residual distribution of sources in $V$ leads to an 
effective surface brightness estimate of $\sim30.3$ mag arcsec$^{-2}$ for the outer disk knots.  
{\it{Right:}} this version only shows weak signal 
above $90\%$ CL and none $>99.9\%$ CL.  In the outermost regions of 
the disk, it is the reddest sources that dominate by number.   
From the original $\sim1600$ sources detected in this radial range,
just $\sim200$ (or $\sim1/8$) remain; accounting for their distribution in $V$ leads to an 
effective surface brightness estimate of $\sim32.1$ mag arcsec$^{-2}$.
\label{Hess_2panel}}
\end{figure*}

Figure~\ref{Hess_2panel} contains background-subtracted 
Hess diagrams of regions between $1.0-1.5\ R_{25}$ and $1.3-1.5\ R_{25}$ 
in the left and right panels, respectively.   
We use low-count, Poisson single-sided upper and lower limits from \cite{Gehrels86} to 
outline signal lying in excess of the background at the $>90\%$ and $>99.9\%$ confidence 
level (CL).  We detect a statistically significant knot population in the outer disk 
of NGC 3184.  The strength of the signal fades quickly 
with radius, and the right panel of Figure~\ref{Hess_2panel} suggests 
that $\sim1.4 R_{25}$ is the limit of how 
far this method can probe the outer disk.  

Although outer disk knots are typically thought of as being young because they were 
discovered in the UV, we find that the population of knots redward of $U-V = -0.2$ 
dominates that of bluer ones.   The younger, blue population of knots is only 
detected near the optical radius with this method.  
However, we cannot distinguish between dust and age as the primary driver of color;  
we assume 
insignificant reddening of the majority of outer disk sources, but caution that 
the relative numbers of young vs.\ old knots is uncertain because of this issue.  

In Figure~\ref{colmag} a $10^4 M_{\sun}$ model star cluster track for ages between 100 Myr and 
1 Gyr provides an envelope for the bright end of the distribution.  Above this track 
the number of objects falls off rapidly, implying that whatever outer disk 
objects are in this sample, the majority have masses $< 10^4 M_{\sun}$.
Our catalog becomes incomplete for objects resembling 
$10^4 M_{\sun}$ clusters older than $\sim1$ Gyr, and the age 
where this incompleteness sets in decreases quickly with decreasing 
cluster mass.
A $100 M_{\sun}$ track suggests that at the low-mass end 
our final catalog is restricted to 
objects younger than a few tens of Myr, although this 
estimate is more uncertain due to the previously-discussed 
stochastic sampling of the IMF.  
Because one massive O-star forms per $200-300 M_{\sun}$ of stars in a cluster 
\citep[][and references therein]{Parker07}, many outer disk clusters 
detected here may host only a few O-stars, and many may contain none at all.

Although not well-constrained with the available optical data (i.e.\ $V$ \& $U$ only), 
our estimated cluster mass range ($\sim10^2 - 10^4 M_{\sun}$) matches that 
found in the extended disk of NGC 4625 \citep[see][their Figure 4]{GildePaz05}.  
Those authors examined GALEX FUV-selected complexes in the 
outer disk of NGC 4625 that also show corresponding H$\alpha$ emission,
and found that the majority of detected 
knots (or clusters) lie between $10^3 - 10^4 M_{\sun}$, with the lowest-mass 
knots resembling $\sim500 M_{\sun}$ clusters.  
A knot containing fifteen $20 M_{\sun}$ O-stars would lie 
just among the range of young ($< 10$ Myr) $\sim10^3 M_{\sun}$ clusters 
\citep[Figure 4,][]{GildePaz05}, 
similar to what we expect to find in our images of the outer disk of NGC 3184.
The distinction between our data and those from GALEX
is that we have superior spatial resolution and in principle can find older clusters.
The strength of the GALEX data relative to ours is that the
young clusters are more prominent relative to background sources.

\section{Clustering of knots in the outer disk}
 
Another way to estimate the outer disk extent is to consider the spatial 
distribution of knots most likely associated with NGC 3184.  Here we 
present correlation functions to trace the self-clustering 
of knots in the outer disk.  To maximize whatever 
correlation signal exists, we must increase the 
contrast between the sources of interest and the background.  
In the FUV and NUV bands of the GALEX images, 
the blue star-forming knots stand out sharply against 
the redder background galaxies.
\cite{Zaritsky07} exploited this color contrast in the GALEX data 
and calculated the two-point correlation function of detected sources 
around nearby galaxies, to quantify the ubiquity of any extended disk 
component. 
Their main result was that five of the eleven galaxies studied 
show an excess of sources 
between $1.25 - 2 R_{25}$, which statistically implies that at least a quarter of such galaxies 
have an extended blue knot population.

We would like to use a statistical method similar to the two-point correlation analysis 
of \cite{Zaritsky07} to measure 
the extent of NGC 3184's outer disk with our LBT data.   The two-point correlation 
approach would work well here if NGC 3184 had large numbers of blue knots in its outer disk, 
because they would create the sharp color and number density 
contrast required to isolate them from the background.  Due to the 
lack of blue outer disk sources, and that the dominant red sources have similar 
$U-V$ colors to the background objects (Figures~\ref{colmag} \& \ref{Hess_2panel}), 
the \cite{Zaritsky07} approach is not effective with our optical data.  
Note that a particular red color cut {\it{can}} be applied to provide 
enough contrast for a significant two-point 
correlation function to be made in the inner disk, 
but this correlation quickly becomes insignificant beyond $R_{25}$.  

We must find another way to accentuate the contrast between knots and 
the background and thereby maximize whatever two-point correlation function signal
may exist in the outer disk.  
As evident in Figure~\ref{knots_fig}, the knots cluster about themselves.  
We use this self-clustering of knots as a means to provide the 
necessary contrast with the background; we assume that the background sources (mostly 
galaxies) do not cluster on scales similar to the disk objects.  
We then ask, `how far can we trace clustered knots?'

\begin{figure}[h]
\centerline{
\includegraphics[angle=0,width=2.0in]{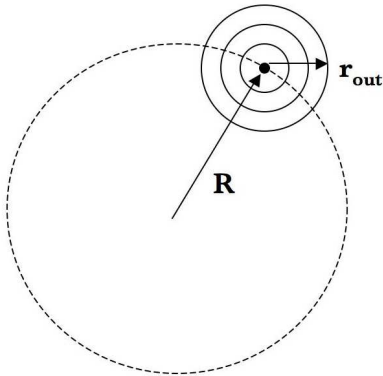}}
\caption{The radii defined for creating the restricted three-point correlation maps.  
\label{radii_expl}}
\end{figure}

To measure the clustering of knots as a function of galactocentric distance, we 
calculate a `restricted' three-point correlation, 
for which one of the 3 points is always defined to be the center of the galaxy.  
We begin by selecting sources by color and magnitude to have rough mass and age constraints. 
We then calculate the two-point knot-knot correlation at each radius. Specifically, 
for each knot we calculate the distance to neighboring knots ($r_{out}$ - see Figure~\ref{radii_expl}), 
bin the distribution of distances, and obtain an estimate of the two-point 
knot-knot correlation function at the galactocentric radius $R$ of this individual knot. 
We repeat this process for every knot and bin in $R$ to obtain the raw
version of the restricted three-point correlation map.

The raw restricted three-point correlation map must be 
corrected for several effects.  Because we are attempting to 
isolate the self-clustering signal of disk-only knots as a function of radius, we 
must first account for contaminating correlation signal in the raw map 
arising from the given distribution of knots in the field at all $R$ (both disk and 
background sources), as well as 
for contaminating signal from any spatially-correlated background galaxies.  
First, we normalize the two-point correlation function 
at each $R$ by dividing by the number of knots at that radius, divide by 
the background number density of sources calculated in an annulus at large $R$, and 
subtract 1 from the result so that the `unclustered' signal is zero.  
Second, we correct for the underlying `self-correlation' of knots across the field 
via Monte-Carlo simulations. 
Self-correlation arises because the knots are correlated with the galaxy, and 
so will appear correlated with each other even if the outer disk has absolutely 
no intrinsic clustering.  To calculate and subtract this we generate 500 three-point 
correlation maps, similar to the original but each time randomizing the knots 
in angle about the galactic center.  
This approach preserves the source density gradient with $R$ 
(which we do not wish to remove) while providing a measure of 
the self-correlation of a disk with no knot clustering. 
Subtracting the average of these 500 realizations then provides a 
measure of the intrinsic knot clustering.
Lastly, we use the two-point correlation function at large $R$ to subtract the underlying 
galaxy-galaxy correlation function (i.e. the apparent clustering of background galaxies). 
If there were no outer disks knots, then the 
correlation function along $r_{out}$ would simply be the galaxy-galaxy correlation 
function and be independent of $R$. By subtracting the $r_{out}$ correlation as 
measured at large $R$ (beyond $2.5 R_{25}$), where there presumably 
are no outer disk knots, we have removed the effect of contaminating galaxies.  
The average amplitude of the random `self-clustering' correction (due to the distribution of 
all sources in the field) is equal to the average amplitude of the raw correlation 
signal at large $r_{out}$ ($>1.5$ kpc), indicating that the knots are not self-clustered 
on large scales.  The average amplitude of the background galaxy-galaxy correction 
is thus also similar to the random-self-clustering-corrected signal at large $r_{out}$, because 
the signal now remaining in these regions is due only to background galaxy-galaxy clustering.  
At smaller $r_{out}$, particularly where we detect statistically-significant signal in the 
final map (described below), the amplitude of this galaxy-galaxy correction 
is negligible compared to the signal remaining from disk objects.

We use the scatter from the random realizations to set the 
significance level of positive excursions in the correlation map.  
Taking the scatter from a `background' region 
of the random maps is insufficient, however.  Due to the source distribution in the 
field and the different annuli area used in constructing 
the maps, the scatter is not uniform.  From our 500 realizations we
identify excursions that have only a 5\% or 1\% chance of occurring
at that particular $R$, $r_{out}$ combination.  
Filled contours connecting these $R$, $r_{out}$ `pixel' excursion values define 
the two-dimensional 95\% and 99\% significance thresholds in the map.  The 
resulting background of noise peaks in the 
map (on scales of the $R - r_{out}$ binsize) can sometimes make it difficult 
to confidently attribute signal to actual knots clustering in the disk.  
In the more obscure cases, clues from the larger structure of the disk may help. 

\begin{figure}[h]
\centerline{
\hspace{1.3in}
\includegraphics[angle=90,width=2.0in]{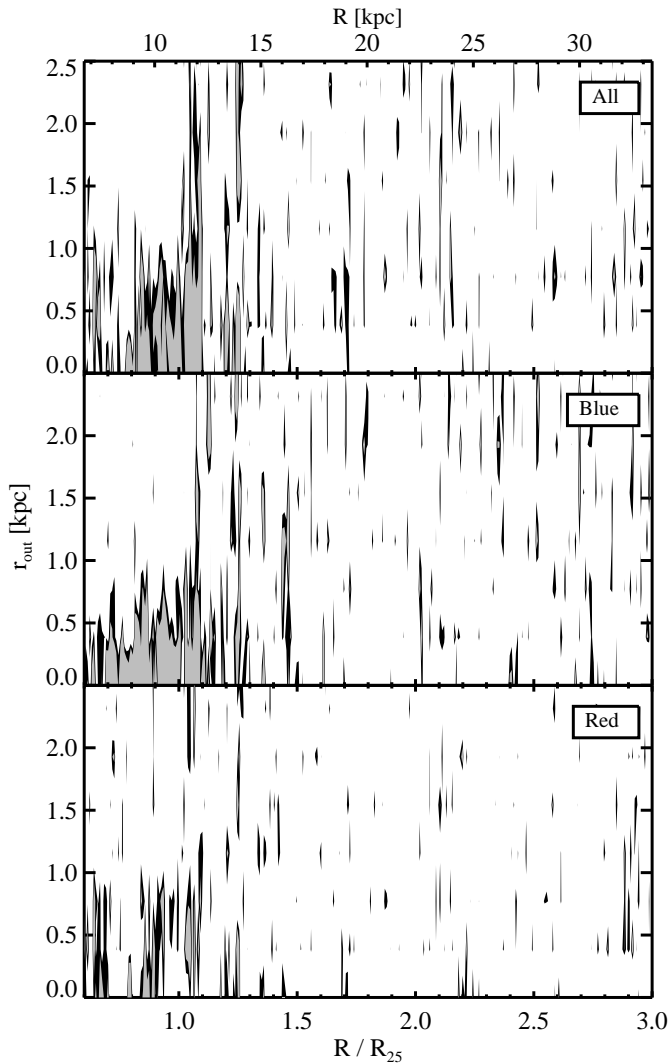}}
\caption{Restricted three-point correlation maps from sources in our final catalog.  
The top panel is from all sources with $-1.7 < U-V < 0.7$ and $18 < V < 27.5$, while the 
lower panels result from splitting the sample into blue and red components on either side of 
$U-V = -0.2$ (middle and bottom panel, respectively).  Black and grey 
show areas where signal is detected at the $>95\%$ and $>99\%$ significance level, 
respectively, as a function of both $R$ and $r_{out}$.  
\label{corr_plot}}
\end{figure}

Figure~\ref{corr_plot} shows the final restricted three-point correlation 
signal above the 95\% and 99\% significance thresholds (in black and 
grey shading, respectively) 
from `all' sources with $-1.7 < U-V < 0.7$ and $18 < V < 27.5$ (top panel), 
and from the sample split into `blue' and `red' on either side of 
$U-V = -0.2$ (middle and bottom panels, respectively).  
If dust significantly reddens outer disk clusters, then 
the strength of the correlation results presented 
will be underestimated for young sources and overestimated for older sources.  

The choice of the red extreme in our $U-V$ color cut 
(0.7) was motivated by two requirements: to 
retain the majority of knots most likely associated with NGC 3184 while 
excluding the maximum number of background sources (mostly galaxies).  
\cite{Fukugita95} show that the $U-V$ colors of 
$z = 0.2$ late-type spirals are near 0.8.  Because these and earlier-type 
galaxies dominate the background, we chose $U-V < 0.7$ as the red extreme, 
with the assumption that this cut retains the majority of knots likely associated with 
the outer disk of NGC 3184 (Figure~\ref{colmag} shows that we will retain the majority 
of sources between $1.0 - 1.5 R_{25}$).  
However, because galaxies become bluer in this set of filters as $z$ 
increases, we expect increasing contamination from more distant galaxies.  
This is exacerbated by the evolution of galaxies because galaxies 
are typically brighter and bluer at higher redshift.  
Using standard $\Lambda$CDM cosmology, the tabulated colors from \cite{Fukugita95} 
and $L$* from \cite{Blanton03}, 
we find that for $V > 23$ $L$* Scd galaxies are included for $z > 0.6$.  
It is therefore not feasible to select a $U-V$ color cut that effectively 
eliminates galaxy contamination to the limits of our photometric sensitivity.  
Because we do 
not exclude all background galaxies with our color cut, we require a 
method to statistically distinguish knots from the background$-$hence the correlation 
map approach.  It is important to note that the red cut is the 
only form of color contrast we use to 
distinguish outer disk sources from the background.

The final restricted three-point correlation maps provide another method 
to estimate the extent of the outer disk 
(the first being the background-subtracted Hess diagrams).  
The blue sources cluster 
uniformly to just beyond the optical radius, and then break into discrete 
aggregates that are detectable out to $\sim1.5 R_{25}$.  
While the red knot correlation signal is noisier, presumably 
due to less color distinction with the background and a less clustered 
distribution, there remain 
regions of significant clustering signal at radii comparable to those found in 
the blue sources.  The red map is made using 
$\sim1.8 \times$ more sources than the blue map.  
Regions of increased signal 
along $r_{out}$ in the blue source map (e.g.\ at 0.71, 0.85, 0.94, and $1.05R_{25}$) 
trace overdensities of knots along the spiral arms seen in the images, as do the 
few regions of increased signal in the red map.  
The regions of increased clustering signal common to both the blue and red maps 
near $0.9R_{25}$ and $1.05R_{25}$ can be visually attributed to large 
aggregates of knots seen in the images.  For example, the overdensity near $1.05R_{25}$ 
is likely influenced by the large aggregate shown in the bottom half of Figure~\ref{knots_fig}.

\begin{figure}[h]
\centerline{
\hspace{1.3in}
\includegraphics[angle=90,width=2.0in]{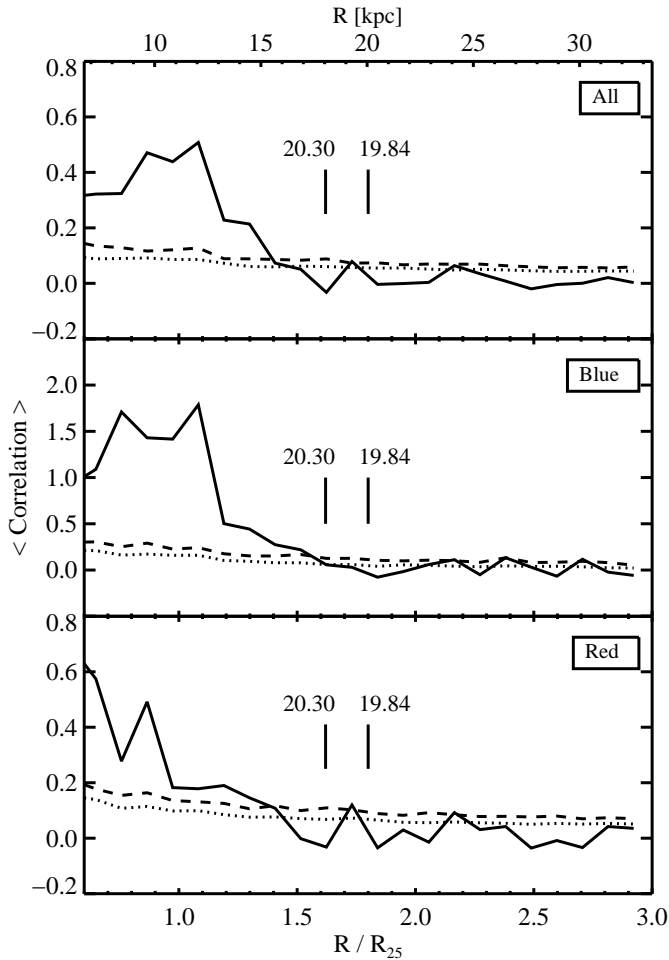}}
\caption{Average correlation signal from areas $r_{out} \leq 0.5$ kpc in 
restricted three-point correlation maps similar to those of Figure~\ref{corr_plot}.  
The top, middle and bottom panels are from the same all, blue and red samples of Figure~\ref{corr_plot}.  
The dotted and dashed lines mark the 95\% and 
$99\%$ significance thresholds, respectively.  Both the blue and red sources 
extend globally out to $\sim1.4 - 1.5 R_{25}$.  Also marked are the approximate 
radial extents of \ion{H}{1} gas for two different boundaries, 
$N(\rm{HI}) > 2\sci{20} \cm{-2}$ and $6.9\sci{19} \cm{-2}$, 
corresponding to the damped Lyman-$\alpha$ (DLA) system threshold 
and the $3\sigma$ threshold of our 
\ion{H}{1} map, respectively.  See \S6 for more details of the \ion{H}{1} data.  The significant 
clustering signal from the blue knots extends nearly to 
the DLA system threshold. 
\label{corr_plot_avg}}
\end{figure}

Due to the irregular signal in Figure~\ref{corr_plot}, it is difficult to 
measure the radial extent of the outer disk directly from these maps.  One way 
to better estimate the radial extent is to average the correlation signal below a 
chosen $r_{out}$ and determine the $R$ at which the average signal falls 
below the local averaged significance thresholds.  When this is 
done using the fine $R$ binning of Figure~\ref{corr_plot}, a correspondingly 
irregular average with $R$ is produced (not shown).  Rebinning the $R$-annuli to 
be $\sim12 \times$ coarser than in Figure~\ref{corr_plot} (to $\sim1.5$ kpc per bin) results in 
Figure~\ref{corr_plot_avg}, which presents a smoother trace of 
average correlation signal in $R$ for $r_{out} \leq 0.5$ kpc.  
Also plotted are the corresponding average 95\% and 99\% significance thresholds in 
dotted and dashed linestyle, 
respectively, made from averaging similarly resampled threshold maps at regions 
$r_{out} \leq 0.5$ kpc to match the data map averaging.  
There is clustering signal in both 
the blue and the red 
knots to $\sim1.4 - 1.5\ R_{25}$.  Also shown in Figure~\ref{corr_plot_avg} 
are the approximate extents of the 
\ion{H}{1} gas disk from two column density thresholds (see the caption and \S6 for details).

Surface brightness measurements of the disk at these radii may be quite difficult.  We defer 
a detailed surface brightness analysis of the diffuse stellar component to an upcoming paper, 
where we will first account for low-level scattered light in the mosaics.  
Here we estimate the {\it{effective}} surface brightness (eSB) of knot aggregates that 
can be detected in our correlation maps.  
We have inserted mock aggregates at different radii to determine the projected 
number density of knots necessary to produce a significant signal in 
our restricted three-point correlation maps.  The mock aggregates are simply 
groupings of `detection' locations added to the field of real knot detection locations; 
once a knot is counted as a detection within a particular color-magnitude cut, 
the shape and color of the knot are irrelevant for the construction of the 
correlation maps--only the locations of the detections are important.  
We therefore make no assumptions of the detailed shapes or colors of the fake knots, 
other than they satisfy the same criteria as the real knots to have made it into the final sample 
being considered.  
We find that a complex of 0.08 knots arcsec$^{-2}$ (eight 
knots randomly distributed within a circular area of diameter $11\farcs2$) at 
$1.4 R_{25}$ would 
be detected as significant signal (above the 99\% significance threshold) 
in a map like those of Figure~\ref{corr_plot}.  
Combining these numbers with the apparent magnitudes from our faintest detected sources 
(27.5 mag in $V$) leads us 
to a limiting $V$-band eSB of $\sim30.2$ mag arcsec$^{-2}$ 
for a significantly clustered aggregate at the limit of our photometric sensitivity.  
This eSB 
is similar to the outer disk eSB found from the $1-1.5 R_{25}$ 
background-subtracted Hess diagram in 
Figure~\ref{Hess_2panel} ($\sim30.3$ mag arcsec$^{-2}$), although 
we caution that these estimates represent different samples of objects 
generated by separate methods.  The eSB 
estimated from the outermost $\sim200$ sources remaining in the 
$1.3-1.5 R_{25}$ background-subtracted Hess diagram 
is $\sim32.1$ mag arcsec$^{-2}$. 
For comparison (primarily to the eSBs estimated from our background-subtracted Hess diagrams), 
$V$-band outer disk eSBs estimated from star counts in M31 
\citep{Irwin05} and NGC 300 \citep{BlandHawthorn05} are $\sim32$ mag arcsec$^{-2}$.  
Low surface brightness isophote fitting is typically limited to $\sim28$ mag arcsec$^{-2}$ 
due to a range of technical challenges \citep{BlandHawthorn05}, and so going fainter 
typically requires a resolved population.  
Our eSBs are from cluster-like knots rather than individual stars, however; 
NGC 3184 is too distant for detecting any but the brightest 
individual stars from the ground.  
Although we have not yet carried out a surface brightness analysis of the 
diffuse outer disk component of NGC 3184, we can estimate the expected surface brightness 
at large radii using results from the literature.  Using \cite{Pompei97} and 
\cite{Godwin77}, we estimate the diffuse stellar component to be 
$\sim28$ mag arcsec$^{-2}$ at $1.5\ R_{25}$, or $\sim2 - 4$ mag brighter than the eSBs 
found from the knots in Figure~\ref{Hess_2panel}.

\section{A comparison to GALEX UV source extent}

\begin{figure}[h]
\centerline{
\includegraphics[angle=90,width=3.5in]{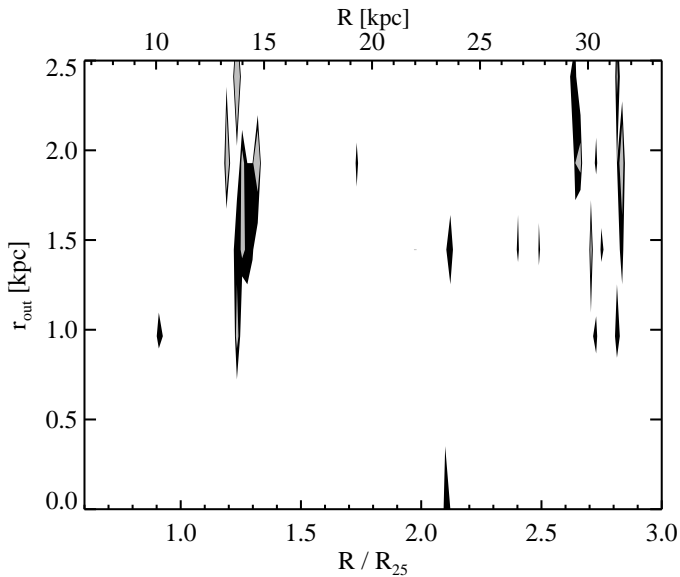}}
\caption{Same as Figure~\ref{corr_plot}, but here using publicly-available 
GALEX AIS sources.
\label{GALEX_corr}}
\end{figure}

Our three-point correlation technique can be easily applied to other 
datasets.  Figure~\ref{GALEX_corr} shows the resulting 
correlation map when considering all GALEX UV sources near 
NGC 3184.  The GALEX sources were collected from a search of the 
Multimission Archive at STScI\footnote{MAST; http://galex.stsci.edu/GR4} which 
provides a catalog of detected sources and their associated photometry.  
Unfortunately the available GALEX data of this 
field is very limited, with only very shallow ($\sim100$ second) exposures 
from the All Sky Imaging Survey (AIS) publicly available.  
Figure~\ref{GALEX_corr} is therefore made from only the brightest 
knots around NGC 3184, corresponding to cluster masses of 
$\sim10^4 M_{\sun}$ or larger.  
Because the majority of our cluster-like detections correspond to 
$M < 10^4 M_{\sun}$ (Figure~\ref{colmag}), these GALEX data 
trace a different cluster population.  
We expect that the correlation map of Figure~\ref{GALEX_corr} 
would reveal structures comparable to those of Figure~\ref{corr_plot} 
with deeper GALEX data, although the resolution of the clustering in $r_{out}$ 
would be lower due to the $\sim6\times$ coarser spatial resolution.  
The correlation signal near $1.3\ R_{25}$ on 
scales $1 < r_{out} < 2$ kpc comes from the 
distribution of knots just visible in the GALEX images 
near the outskirts of the disk, seemingly associated 
with extended spiral arms.  The detections decline in the inner disk, presumably 
because it is more difficult to detect individual knots over the bright extended emission.

\section{A comparison to the underlying neutral gas profile}

\begin{figure}[h]
\centerline{
\includegraphics[angle=0,width=3.4in]{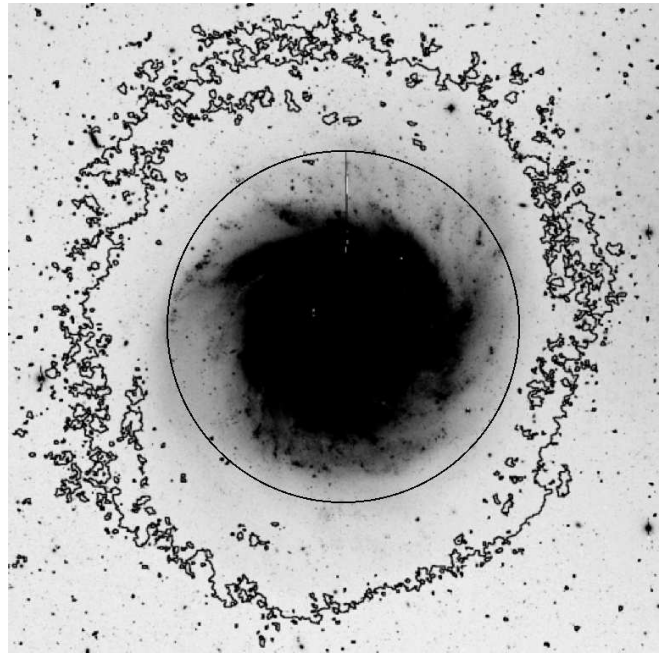}}
\caption{The central portion of our deep LBT/LBC $V$-band mosaic 
with the $3\sigma\ N(\rm{HI})$ threshold in black contour.
This image is $\sim3.7 R_{25}$ or $\sim12.9'$ on a side (the black 
circle has a radius of $R_{25}$).   
The gas disk extends beyond our detections of 
the outermost regions of the optical disk, as 
probed by our background-subtracted Hess diagrams and restricted 
three-point correlation maps.
\label{image_HI}}
\end{figure}

\begin{figure}[h]
\vspace{0.4cm}
\centerline{
\includegraphics[angle=90,width=3.5in]{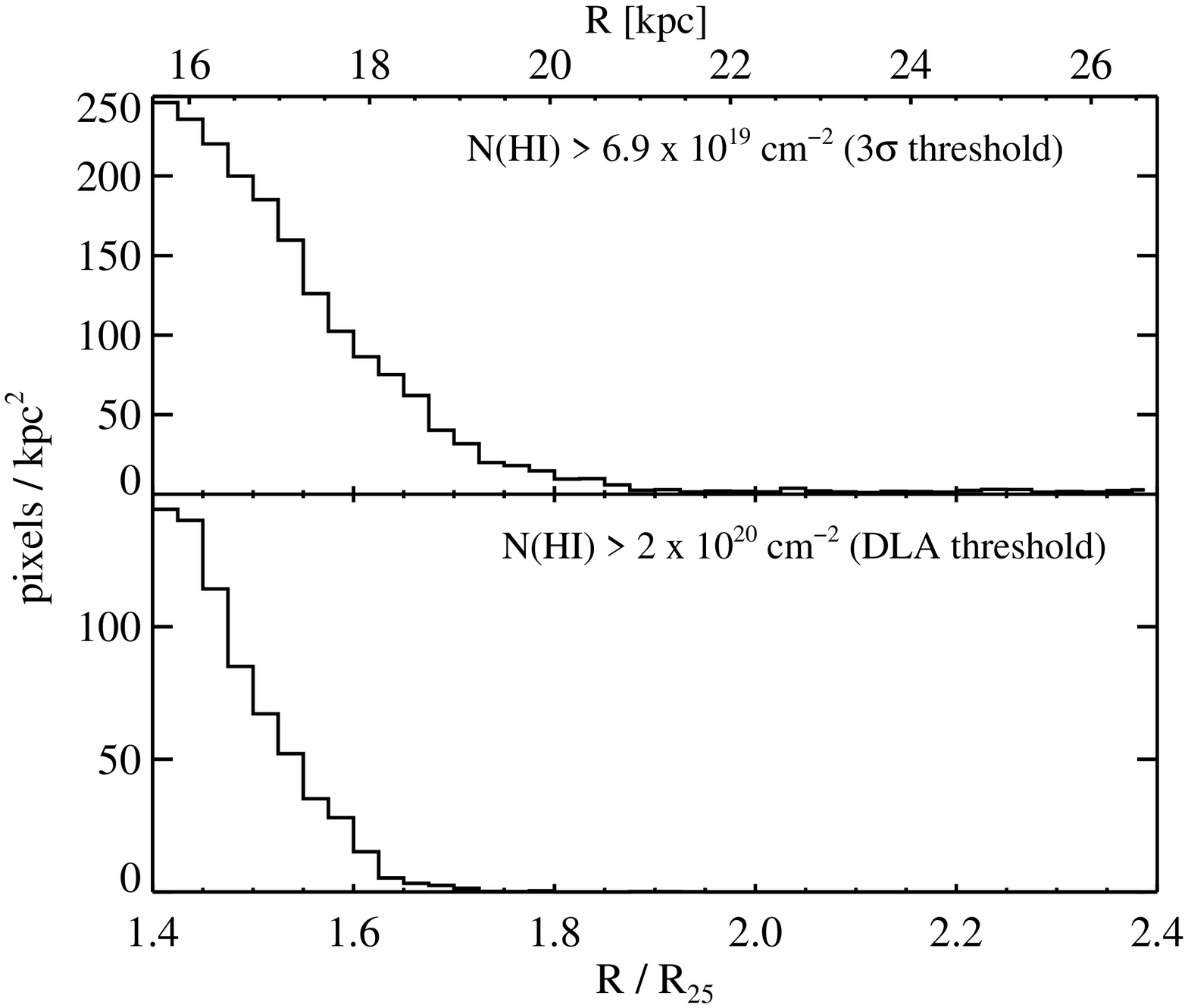}}
\caption{Histograms showing the number of \ion{H}{1} pixels with 
$N(\rm{HI}) > 6.9\sci{19} \cm{-2}$ (top) 
and $N(\rm{HI}) > 2\sci{20} \cm{-2}$ (bottom) per kpc$^2$ in circular 
annuli of width $0.025 R_{25}$.  We 
estimate the radial extent of the \ion{H}{1} gas above the $3\sigma$ and DLA thresholds to be 
$1.80 R_{25}$ and $1.62 R_{25}$, respectively.  
These extents were marked on Figure~\ref{corr_plot_avg} to compare with 
the extent of clustered knots from the correlation map approach.
\label{HI_extent}}
\end{figure}

We now compare our estimates of the stellar/cluster disk extent 
to the size of the underlying gaseous disk.  
We begin by measuring the extent of the neutral gas disk from 
existing 21cm Very Large Array (VLA) imaging 
of NGC 3184.  The beamsize of the C-configuration 
data is $10''$ with natural weighting (each visibility is weighted by the inverse of the 
noise variance, giving more weight to shorter baselines and larger spatial scales).  
The Astronomical Image Processing System\footnote{AIPS; http://www.aips.nrao.edu/} 
software package was used to reduce the data.  
Figure~\ref{image_HI} shows the $3\sigma\ N({\rm{HI}})$ extent 
($N(\rm{HI}) = 6.9\sci{19} \cm{-2}$) overlaid on a portion of our deep LBT $V$-band mosaic.  
To quantify the \ion{H}{1} radial extent, 
Figure~\ref{HI_extent} presents histograms showing the number of \ion{H}{1} pixels per kpc$^2$ 
above two 
$N(\rm{HI})$ threshold values, as a function of $R$ in $0.025 R_{25}$-wide annuli (chosen 
for a smooth, well-sampled trace of the decline).  
We use circular annuli for convenience; 
the inclination of NGC 3184 is $17^\circ$ \citep{Daigle06}, or nearly face-on.  
The top panel is of pixels with $N(\rm{HI}) > 6.9\sci{19} \cm{-2}$ (the $3\sigma$ threshold and 
contour shown in Figure~\ref{image_HI}), and the bottom 
panel is of pixels with $N(\rm{HI}) > 2\sci{20} \cm{-2}$.  The latter value 
was chosen to match the damped Lyman-$\alpha$ (DLA) system threshold, which 
distinguishes between predominantly neutral and predominantly 
ionized gas \citep[see][and references therein]{Wolfe05}.  
The $N({\rm{HI}}) = 2\sci{20} \cm{-2}$ boundary represents the edge 
of the dominant reservoir of neutral gas available for star formation in the disk.  
We estimate the gas disk extent to be $1.62 R_{25}$ and $1.80 R_{25}$ for 
the DLA and $3\sigma$ thresholds, respectively.  
Comparing these extents to those of the knots found using 
our Hess diagrams and three-point correlation maps ($\sim1.4-1.5 R_{25}$; 
Figures~\ref{Hess_2panel} \& \ref{corr_plot_avg}), we see that the 
optical disk extends at least to just below the DLA system threshold, but 
that the full $3\sigma\ N(\rm{HI})$ gas disk extends to larger radii.

\begin{figure}[h]
\centerline{
\hspace{1.3in}
\includegraphics[angle=90,width=2.0in]{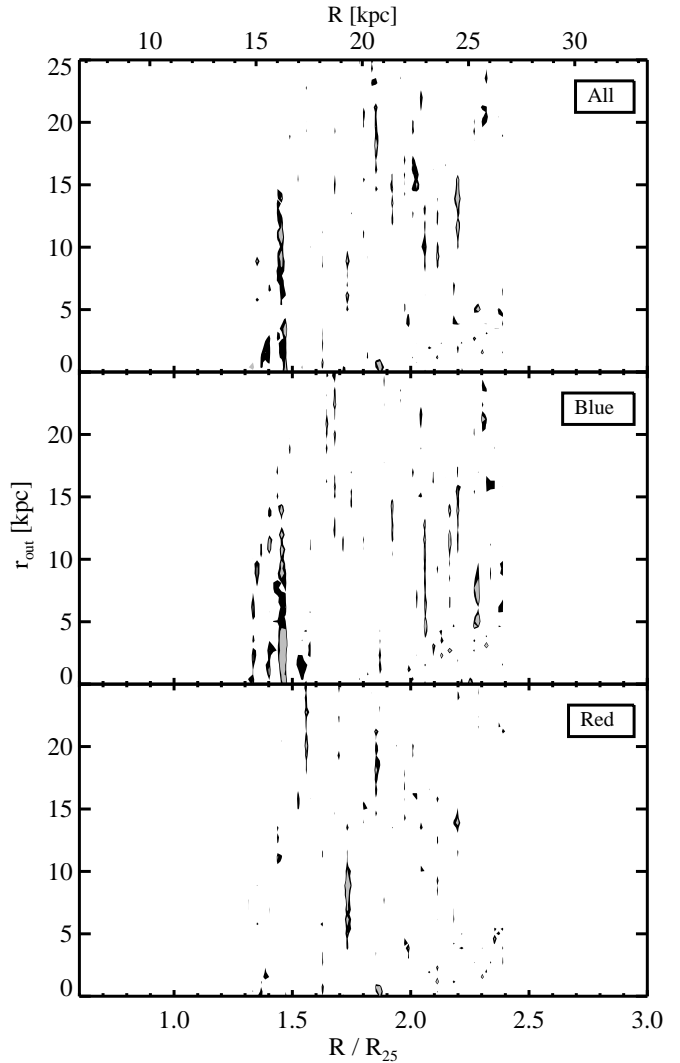}}
\caption{Restricted three-point cross-correlation maps of 
LBT-detected knots and \ion{H}{1} pixels with $N(\rm{HI}) > 6.9\sci{19} \cm{-2}$.  
The top panel is from all sources with $-1.7 < U-V < 0.7$ and $18 < V < 27.5$, while the 
lower panels result from splitting the sample into blue and red components on either side of 
$U-V = -0.2$ (middle and bottom panel, respectively). 
Only data between $1.3-2.4 R_{25}$ were used.  
Black and grey 
show areas where signal is detected at $>95\%$ and $>99\%$ significance, respectively.
\label{LBT_HI_corr}}
\end{figure}

The degree to which the extended cluster and gas disks are related can 
be quantified by the cross-correlation of LBT-detected knots and detected 
\ion{H}{1} signal.  This cross-correlation can be used as another method to trace 
the extent of the optical disk, in the case where the knots do not cluster about 
themselves but rather with the \ion{H}{1} gas.  The latter would be expected 
(on small $r_{out}$ scales) if the knots 
formed directly from the gas in the extended disk (discussed further in $\S8$).  
We construct restricted three-point 
cross-correlation maps in a similar manner as the 
knot-knot correlation maps but instead using \ion{H}{1} pixels above the 
$3\sigma\ N(\rm{HI})$ threshold ($6.9\sci{19} \cm{-2}$; 
Figure~\ref{image_HI}) as the objects surrounding a 
particular LBT-detected knot.  We also do not subtract a background 
galaxy correlation signal from the maps because optical background galaxies are 
not correlated with the 21cm peaks in the background of the \ion{H}{1} map.  
The knot-knot clustering analysis 
showed that the optical disk extends to at least 1.4$R_{25}$, so we have restricted the 
lower boundary of radii considered here to 1.3$R_{25}$ 
(\ion{H}{1} and knot data below 1.3$R_{25}$ were ignored when making these maps).  
The upper range of the radii considered here 
is 2.4$R_{25}$ due to the extent of available \ion{H}{1} data.   The upper, middle and 
lower panels of Figure~\ref{LBT_HI_corr} show the knot-\ion{H}{1} pixel 
cross-correlation signal from the all, blue and red samples, respectively.  
We remind the reader that spurious `significant' signal is to be expected across the map at the $1 - 5 \%$ 
level, given the large number of resolution elements in the map, and so it is sometimes difficult 
to know if a particular signal in a map is real.  Features viewed with skepticism, especially those 
at large $R$ or $r_{out}$, or those seen in the map of one sample (blue or red) but not 
in the all map, should be confirmed or refuted with further observations.  
Here we find strong blue knot-\ion{H}{1} pixel clustering signal at 1.45$R_{25}$ extending to 
$r_{out} \sim 15$ kpc, likely due to knots distributed along a tightly 
wound \ion{H}{1} spiral arm seen in the \ion{H}{1} image at $\sim1.45R_{25}$.  The red knot-\ion{H}{1} 
pixel cross-correlation map shows significant extended 
signal near 1.73$R_{25}$ between 
5-10 kpc in $r_{out}$.  This signal likely results from knots lying along another 
\ion{H}{1} spiral arm near the diffuse edge of the \ion{H}{1} pixels considered 
here (see Figure~\ref{image_HI} and 
\ref{HI_extent}); it is however surprising that the red knots are the ones that correlate at 
these distances with the \ion{H}{1}.  
Assuming that the cross-correlation signal at large $R$ in the red map is real, 
the results of Figure~\ref{LBT_HI_corr} show that we can trace knots 
farther out with the knot-\ion{H}{1} cross-correlation than with the knot-knot correlation 
presented in $\S4$.  While correlations with the \ion{H}{1} on these scales are unlikely to 
result from in situ formation of the knots, this cross-correlation 
technique still provides an estimate of the radial extent of 
knots tracing extended spiral structure in the gas.

\section{The local environment of NGC 3184}

There have been suggestions that star formation in outer disks is 
triggered by galactic interactions.  Two of the most 
extended UV disks discovered by GALEX \citep[M83 and NGC 4625, both with UV 
knots extending to four times their optical radii;][]{Thilker05, GildePaz05} are both from 
interacting systems.  Here we search the local environment of NGC 3184 for any possible 
perturbers.  

\cite{Tully88} has shown that NGC 3184 belongs to a small association 
of galaxies consisting of three others (NGC 3104, NGC 3198 and NGC 3319).  
Although our optical and \ion{H}{1} images show no obvious morphological 
signs of disturbance (perhaps some low-level lopsidedness in the gaseous component, 
see Figure~\ref{image_HI}), the H$\alpha$ velocity field of NGC 3184 from \cite{Daigle06} 
does suggest kinematic anomalies, particularly towards the outer regions.  
Using the criteria of \cite{Bailin08} to select candidate satellite 
galaxies most likely associated with a host, we 
query NED\footnote{http://nedwww.ipac.caltech.edu/} for 
extragalactic objects within 700 kpc of NGC 3184 (at the distance of NGC 3184, 
or within $\sim215'$) and with $\delv < 750$ \kms and 
find ten candidate satellites near NGC 3184.  Seven of these ten 
appear to be very small and faint in SDSS-DR5 images, however, and one of the 
remaining three is NGC 3104, a member of the galaxy association mentioned above.  
Therefore, although NGC 3184 does not appear to be interacting or 
have had significant interactions recently (given the lack of obvious morphological 
disturbance), we cannot discount the possibility that previous minor perturbations have 
influenced the buildup of this stellar outer disk. 

However, we add a final word of caution to the suggestion that interactions are required for 
triggering visible outer disks.  \cite{Christlein08} detect extended H$\alpha$ emission from 
a sample of nearby (edge-on) galaxies selected against nearby companions 
and any morphological  disturbances.  These results suggest that visible outer disks are a 
natural result of disk galaxy evolution and do not require triggering by external sources 
(although external disturbances may still raise the level of star formation above 
the nominal value).

\section{Summary and Discussion}

We present initial results from our optical 
extended disk imaging study using the LBC on the LBT.  We will 
present a similar analysis of our full sample (nine galaxies) in a subsequent paper.  
Here we focus on the large, face-on galaxy NGC 3184. 
We find that the outer disk is populated by 
marginally-resolved cluster-like objects (or `knots') with 
masses $\sim10^2 - 10^4 M_{\sun}$ and ages up to $\sim1$ Gyr.  The cluster masses 
and ages are not well-constrained with the available optical data and are meant 
as illustrative estimates only.  
Background-subtracted Hess diagrams show statistically significant numbers of 
these cluster-like objects extending to $\sim1.4 R_{25}$, with the redder knots 
extending to the largest radii ($\sim1.4 - 1.5 R_{25}$).  
We construct restricted three-point 
correlation maps to measure the self-clustering of the detected knots and find significant correlation 
signal extending to $\sim1.5 R_{25}$.  The effective surface brightness of the outer disk 
cluster populations we detect ranges from $30 - 32$ mag arcsec$^{-2}$ in 
V, depending on the method and radii used for the estimate (the restricted 
three-point correlation map approach or background-subtracted Hess diagrams, 
respectively).  We also present a knot-knot correlation 
map from GALEX-detected objects near NGC 3184 and find significant correlation 
near $\sim1.2 R_{25}$, although those data are currently very limited 
(the publically-available GALEX imaging of this field is very shallow and does not 
provide a large number of knots to use in the correlation analysis).  
We compare the 21cm \ion{H}{1} gas disk extent to that of the knots 
and find that the correlated clusters extend nearly to 
the DLA threshold of \ion{H}{1} gas ($2\sci{20} \cm{-2}$; $1.62 R_{25}$) and are
well contained within the $3\sigma$ extent of the \ion{H}{1} gas ($6.9\sci{19} \cm{-2}$; $1.80 R_{25}$).  
The cross-correlation between optical knots and \ion{H}{1} pixels above the $3\sigma$ 
threshold suggests that the blue knots are correlated with \ion{H}{1} spiral 
structure to $\sim1.45R_{25}$, and that the red knots may be correlated with 
the outer fringes of the \ion{H}{1} disk to $\sim1.73R_{25}$.  The radii at which we detect 
significant cross-correlation signal correspond to where we observe obvious spiral structure 
in the HI map, as well as a (less-obvious) enhancement of cluster aggregates tracing the spiral pattern.  
We therefore expect that these structures are responsible for the positive signal.  
Incorporating the \ion{H}{1} data further increases our ability to detect knots 
associated with the disk, to the largest radii.  The cluster-like objects appear to 
extend to the outermost reaches of the gas disk.

Assuming that knots in NGC 3184 formed in clustered aggregates of similar scales over 
the past few Gyr, the fact that we see continued clumpiness 
in the red panel of Figure~\ref{corr_plot} 
suggests that at least some of the original aggregates survive as a group, 
even after hundreds of Myr.  The aggregates and knots that do dissolve, 
perhaps via stellar evaporation caused by encounters with neighboring clusters 
or giant molecular clouds (albeit on longer timescales and with lower efficiency 
than expected in the inner disk), 
may distribute many of their stars into the diffuse, 
low surface brightness component visible in the images (mainly as 
faint, extended spiral structure).  This low surface brightness component 
is better seen in the $V$-band data, as expected by the numbers of red to blue clusters, 
the assumption that the larger ages of red sources means they have had more 
time to evaporate stars into a diffuse component, and simply because the $V$ image 
is deeper than the $U$ image.

The restricted three-point correlation results 
illustrate how even a galaxy like NGC 3184, which appears to lack 
any obvious young outer disk 
component in both our deep exposures and the available GALEX data (albeit limited), 
still hosts cluster populations in its outer reaches.  
This suggests that the \cite{Zaritsky07} and \cite{Thilker07} estimates of the 
local extended UV disk fraction ($\sim$30\%) may lie far below 
the local {\it{optical}} extended disk fraction.

Whether the stellar populations in the outer disk were born in situ 
or scattered from smaller radii is still undetermined.  
\cite{Roskar08a,Roskar08b} have convincingly demonstrated that stars will,
over time, be scattered to larger radii and populate an outer disk of characteristics 
consistent with those measured from unresolved stellar populations 
\citep[e.g.][]{Pohlen02}. However, various features of our knot population 
suggest that these were born in situ.  First, we find knot-knot clustering out to 
interknot separations of 1 kpc.  A scattering model would presumably populate 
the outer disks in a much more stochastic manner. Scattering velocities would 
have to be coordinated to better than 10\% to maintain coherence over 1 kpc 
while scattering objects 10 kpc from their birthsite.  Second, we find 
correlation between the knots and the \ion{H}{1}. The \ion{H}{1} is clearly in a stable rotating 
disk and could not be scattered from smaller radii. The correlations between 
the two suggest a causal relationship.

A recent simulation by \cite{Bush08} presents 
an in situ formation picture for the knots, showing how 
local overdensities of gas in an outer disk can lead to star formation 
closely resembling the UV knots observed by GALEX.  
Furthermore, \cite{Christlein08} detect very regular outer disk rotation curves 
(i.e.\ flat, low velocity dispersion) from deep H$\alpha$ 
spectroscopy of many nearby edge-on disks, suggesting that outer 
disks are merely less-populated extensions of their inner counterparts 
(and therefore that in situ cluster formation should be expected).  
Given the successes of the 
\cite{Roskar08a,Roskar08b} studies, as well as that of \cite{Bush08}, 
we expect both radial migration and in situ formation 
to significantly influence the populations of outer disks.  
We encourage the modelers to identify key signatures of each mechanism 
that can be used to further discriminate between these possibilities.  
With the identification of large 
populations of knots, correlation statistics appears to be a promising 
avenue for such efforts.

\vspace{0.2in}
We would like to thank John M. Hill and 
Olga Kuhn for their efforts during the LBT observations, 
as well as David Sand and Vincenzo Testa 
for helpful discussions concerning LBC data reduction and mosaicing techniques.  
We also thank Daniel Eisenstein for an illuminating discussion of our restricted 
three-point correlation map approach, and Benjamin Weiner for other helpful 
discussions.  DZ and SHF were partially supported under 
NASA LTSA NNG05GE82G and NSF AST-0307482.  
JM acknowledges support from NASA grant 06-GALEX06-0030
and Spitzer grant G05-AR-50443.

\end{document}